\documentclass[12pt,preprint]{aastex}

\shorttitle{Albus~1}
\shortauthors{J. A. Caballero \& E. Solano}

\begin{document}
\title{Albus~1: A very bright white dwarf candidate}

\author{Jos\'e Antonio Caballero\altaffilmark{1,2}}
\affil{Max-Planck-Institut f\"ur Astronomie, K\"onigstuhl 17, D-69117
Heidelberg, Germany}
\email{caballero@mpia.de}
\author{Enrique Solano\altaffilmark{2}}
\affil{Laboratorio de Astrof\'{\i}sica Espacial y F\'{\i}sica Fundamental, INSA,
P.O. Box 50727, E-28080 Madrid, Spain}
\email{esm@laeff.inta.es}

\altaffiltext{1}{Alexander von Humboldt Fellow at the MPIA.}
\altaffiltext{2}{Members of the Spanish Virtual Observatory Thematic Network.}

\begin{abstract}
We have serendipitously discovered a previously-unknown, bright source ($B_T$ =
11.75$\pm$0.07\,mag) with a very blue $V_T-K_{\rm s}$ color, to which we have
named Albus~1. 
A photometric and astrometric study using Virtual Observatory tools has shown
that it possesses an appreciable proper motion and magnitudes and colors 
very similar to those of the well known white dwarf G~191--B2B.
We consider Albus~1 as a DA-type white dwarf located at about 40\,pc.
If confirmed its nature, Albus~1 would be the sixth brightest isolated white
dwarf in the sky, which would make it an excellent spectrophotometric standard. 
\end{abstract}


\section{Introduction}
\label{intro}

The three classical white dwarfs were, at the beginning of the twentieth
century, \object{$o^2$~Eri~B}, \object{Sirius~B}, and the
\object{van~Maanen's~star}.
Although $o^2$~Eri~B had been discovered by Herschel (1785) and Sirius~B had
been predicted by Bessel (1844), it was not until the 1920s when other great
astronomers noticed their oddness (Luyten 1922) and popularized the term ``white
dwarf'' (Eddington 1924). 
The extreme physical conditions by which the white dwarfs are supported against
gravitational collapse could not be understood until the Quantum Mechanic was
properly developed. 
Since the 1930s, the number of known white dwarfs has exponentially increased,
from 18 in 1939 (Schatzman 1958) and over a hundred in 1950 (Luyten 1950) to a
few thousands at present (McCook \& Sion 1999; Eisenstein et al. 2006).
See Liebert (1980), Koester (2002) and Hansen \& Liebert (2003) for extensive
reviews of the general properties of white~dwarfs.

The vast majority of the known white dwarfs are very faint, with typical
magnitudes in the optical from $V$ = 15 to 20\,mag, or even fainter.
Only a few very bright white dwarfs ($V <$ 12\,mag), including the three
classical white dwarfs and \object{Procyon~B}, are known.
Many of them, especially those that are not in double degenerate systems, are
extensively used as spectrophotometric stars (see, e.g., the recent catalogue by
Landolt \& Uomoto 2007). 
Except for rare exceptions, as the very hot dwarfs and cataclysmic variables
found in the {\em ROSAT} all-sky survey of extreme-ultraviolet sources by Pounds
et al. (1993), all the very bright white dwarfs were discovered during
photometric and astrometric surveys before the early 70s 
(e.g. Kuiper 1941; Luyten 1949; Thackeray 1961; Eggen \& Greenstein 1965;
Giclas, Burnham \& Thomas 1965; Schwartz 1972).
Afterwards, and in particular with the advent of the Sloan Digital Sky Survey, 
the detection of new white dwarfs and blue subdwarfs has been biased towards
magnitudes fainter than $V$ = 12\,mag.
Because of that, the photometry-based discovery of a very bright white dwarf
candidate 35 years later would be outstanding.
If confirmed, it would yield doubts on the real knowledge that we have of the solar
neighbourhood and on the completeness of previous and current surveys for
white~dwarfs. 

\begin{figure}
\plotone{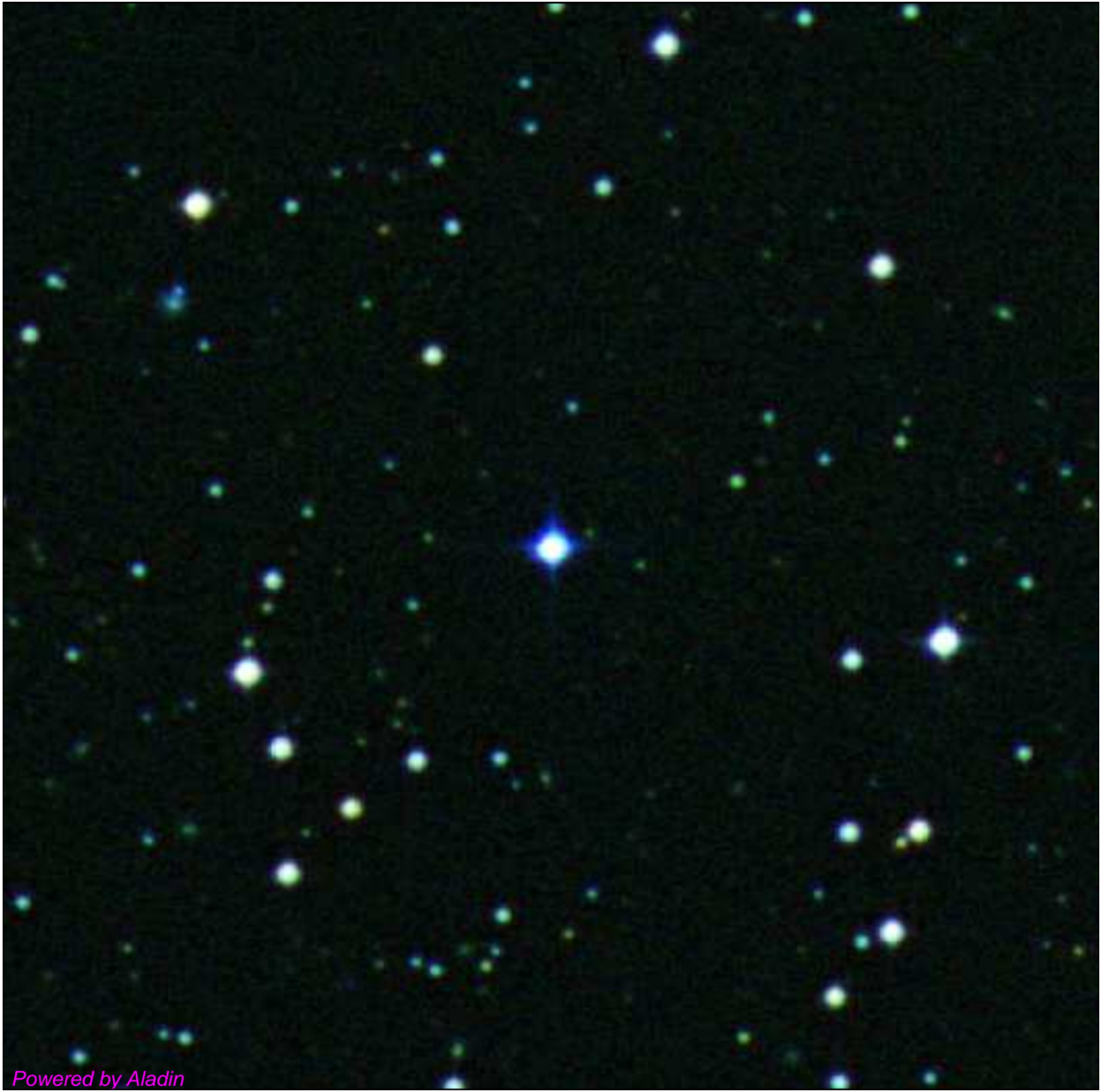}
\caption{False-color composite image, 5.6\,$\times$\,5.6\,arcmin$^2$ wide,
centred on Albus~1. 
Blue is for $B_J$, green for $R$, and red is for $I_N$ (DSS1 and DSS2
photographic plates from ESO and MAMA).
North is up, east is~left.
\label{findingchart}}
\end{figure}

\begin{figure}
\plotone{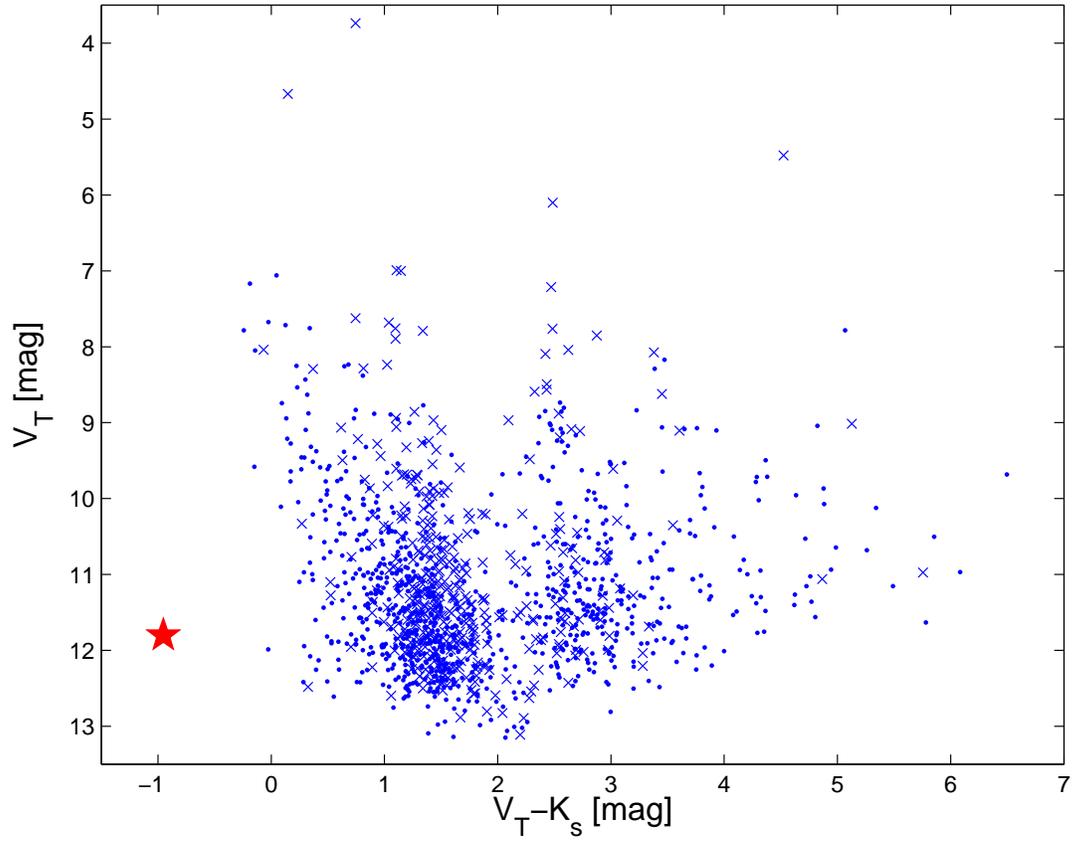}
\caption{$V_T$ vs. $V_T-K_{\rm s}$ color-magnitude diagram
from the data in Caballero \& Solano (2007). 
Tycho-2/2MASS sources with proper motions larger and smaller than
15\,mas\,a$^{-1}$ are shown with crosses and dots, respectively. 
Albus~1 is highlighted with a big filled (red) star.
\label{colormagnitude}}
\end{figure}

In this work we present \object{Albus~1}, a previously-unknown very bright
($V_T$ = 11.80$\pm$0.14\,mag) white dwarf candidate\footnote{{\em Albus} is the
Latin term for~``white''.}.
Its finding chart is provided in~Fig.~\ref{findingchart}.
Albus~1 was serendipitously discovered during an optical-near
infrared photometric study by Caballero \& Solano (2007), devoted to
characterize the young stars and brown dwarfs surrounding \object{Alnilam}
($\epsilon$~Ori) and \object{Mintaka} ($\delta$~Ori). 
As part of this study, they made a correlation between the Tycho-2 (H{\o}g et
al. 2000) and the 2MASS (Cutri et al. 2003) catalogues in ten 45\,arcmin-radius
comparison fields at the same galactic latitude of the brightest stars of the
young \object{Ori~OB~1~b} Association (the Orion Belt; $b \sim$ --17.5\,deg).
The total investigated area was only 17.7\,deg$^2$ ($\sim$0.04\,\% of the whole
sky). 
Albus~1 has a $V_T-K_{\rm s}$ color that clearly deviates from those of the
other 1275 investigated sources (see Fig.~\ref{colormagnitude}).
In particular, while the bluest remaining sources have $V_T-K_{\rm s} \gtrsim$
--0.3\,mag, Albus~1 has a color $V_T-K_{\rm s}$ = --0.95$\pm$0.14\,mag.
The Tycho-2 $B_T V_T$ and 2MASS $JHK_{\rm s}$ photometry shows that the object
is extremely blue at all the wavelenghts from 0.4 to 2.2\,$\mu$m.
Given the extreme blueing of Albus~1, we decided to investigate it in~detail.

\section{Analysis}
\label{analysis}

In this work we have taken advantage of the tools offered by the Virtual
Observatory (VO; {\tt http://www.ivoa.net}), which is an international,
community-based initiative to provide seamless access to the data available from
astronomical archive and services.  
The VO also aims to provide state-of-the-art tools for the efficient analysis of
this huge amount of information. 
In particular, we have used
Aladin ({\tt http://aladin.u-strasbg.fr/aladin.gml}), a VO-compliant  
interactive sky atlas developed by CDS that allows the user to visualize 
and analyze astronomical images, spectra and catalogues available from 
the VO~services.

Albus~1 has an appreciable Tycho-2 proper motion of 19\,mas\,a$^{-1}$.
Since the comparison fields are relatively close to the antapex (the point the
Sun is moving away from), the foreground objects in this region have not very
large proper motions except for some very nearby stars with large tangential
velocities. 
Indeed, the proper motion of Albus~1 is in the percentile 23\,\% of the
investigated sources (i.e. 
77\,\% of the investigated Tycho-2 stars have $\mu <$ 19\,mas\,a$^{-1}$). 
This makes Albus~1 to be a nearby Galactic~object.

Apart from the coordinates, proper motion and $B_T V_T$ magnitudes from Tycho-2
and the $JHK_{\rm s}$ magnitudes from 2MASS, taken from Caballero \& Solano
(2007), we have also collected additional photometric and astrometric data from
other catalogues: SuperCOSMOS Science Archive (Hambly et al. 2001), USNO-B1
(Monet et al. 2003), NOMAD1 (Zacharias et al. 2005) and DENIS (DENIS Consortium
2005).  
They nicely match between them and Tycho-2 and 2MASS, except for the fact that
for very blue objects the Tycho-2 $B_T$ and photographic $B_J$ photometry are
not comparable.
To avoid superfluous, repetitive information in Table~\ref{albus1}, we provide
only the Tycho-2 and 2MASS information and the $RI_N$ photometric data
from~USNO-B1.   

The search with Aladin concluded that there is no radio (NRAO VLA), mid-infrared
($IRAS$), ultraviolet ($EUVE$), X-ray ($ROSAT$) source, or object discussed in
the literature (SIMBAD) at less than 4\,arcmin to Albus~1. 
Neither spectroscopic information exists nor photometry in the Johnson 
passbands has been obtained~yet.

\section{Results}
\label{results}

\begin{figure}
\plotone{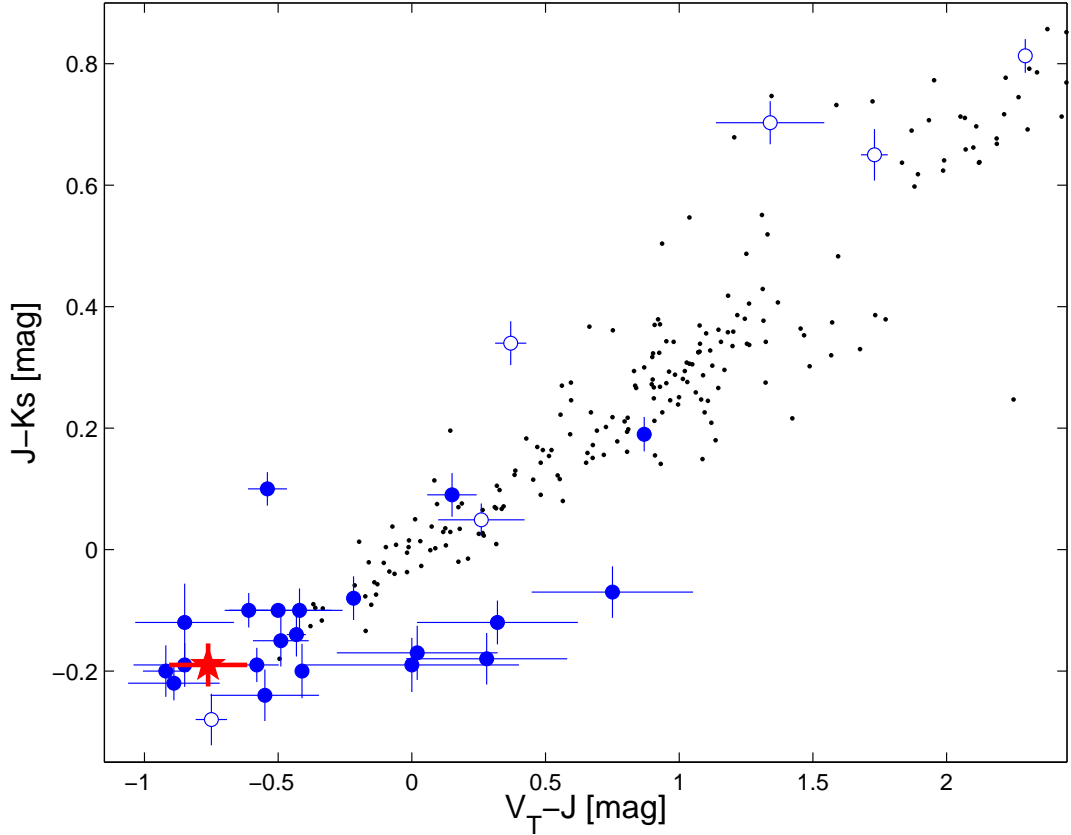}
\caption{$V_T-J$ vs. $J-K_{\rm s}$ color-color diagram of the white
dwarfs and blue subdwarfs listed in Table~\ref{WDs}. 
Code: 
(red) filled star, Albus~1;
(blue) filled circles, single white dwarfs; 
(blue) open circles, white dwarfs in unresolved systems or hot subdwarfs. 
GJ~169.1~B and V841~Ara, that were resolved by Tycho-2 but not by 2MASS, are not
shown; 
(black) dots, a sample of dwarf and giant stars covering the whole spectral type
range from late O to early M, taken from Caballero \& Solano~(2007).
\label{colorcolor}}
\end{figure}

To ascertain the real nature of Albus~1, we must compare its photometry with
that of other very blue objects. 
There is a limited number of Galactic objects with $V_T-K_{\rm s}$ colors as
blue as those of Albus~1: white dwarfs, hot subdwarfs, and early-type main
sequence, blue horizontal branch and Population~II stars.
Fig.~\ref{colorcolor} compares the optical-near infrared colors of Albus~1 with
those of dwarf and giant stars in the direction to Alnilam and Mintaka.
Our blue source is even bluer than the late O- and early B-type stars.
Besides, such luminous stars are located at long heliocentric distances
($d \gtrsim$ 0.4\,kpc), which implies very low proper motions, in contrast with
what we have measured for Albus~1.
Population~II stars are common in the bulge near the centre of the Galaxy and in
the Galactic halo. 
Some of the latters cross the Solar neighbourhood, but only a few of them
display extremely blue colors (see, e.g., the recent photometric study of
horizontal-branch and metal-poor candidates by Beers et al. 2007). 
Therefore, Albus~1 is an early-type hot subdwarf or a white~dwarf.

The extreme blueing of Albus~1 prevents from transforming the $B_T V_T$
magnitudes to Johnson $BV$ magnitudes using standard relations and, therefore,
to compare its colors with other white dwarfs tabulated in exhaustive works such
as in Bergeron, Leggett \& Ruiz (2001).
A new comparison may come, instead, from the available Tycho-2 and 2MASS data.
In Table~\ref{WDs} we have compiled the basic data of the brightest white dwarfs
and blue subdwarfs identified in the Tycho-2 catalogue.
We used the white dwarf lists by McCook \& Sion (1999) and Holberg, Oswalt \&
Sion (2002), and looked for the Tycho-2 counterparts of the white dwarfs
brighter than $V$ = 13.0\,mag.
Fainter objects were not considered due to their poor photometric accuracy. 
This list surpasses the sample of white dwarf observed by the $Hipparcos$
satellite in Vauclair et al. (1997).
There are three evident absences: Sirius~B, $o^2$~Eri~B, and Procyon~B (the
three brightest known white dwarfs), which are too close to other bright stars
and were, thus, not identified by Tycho-2.  
Among the tabulated objects, there is only one hot subdwarf with blue colors,
GJ~3435, which indicates its~rarity.

\begin{figure}
\plotone{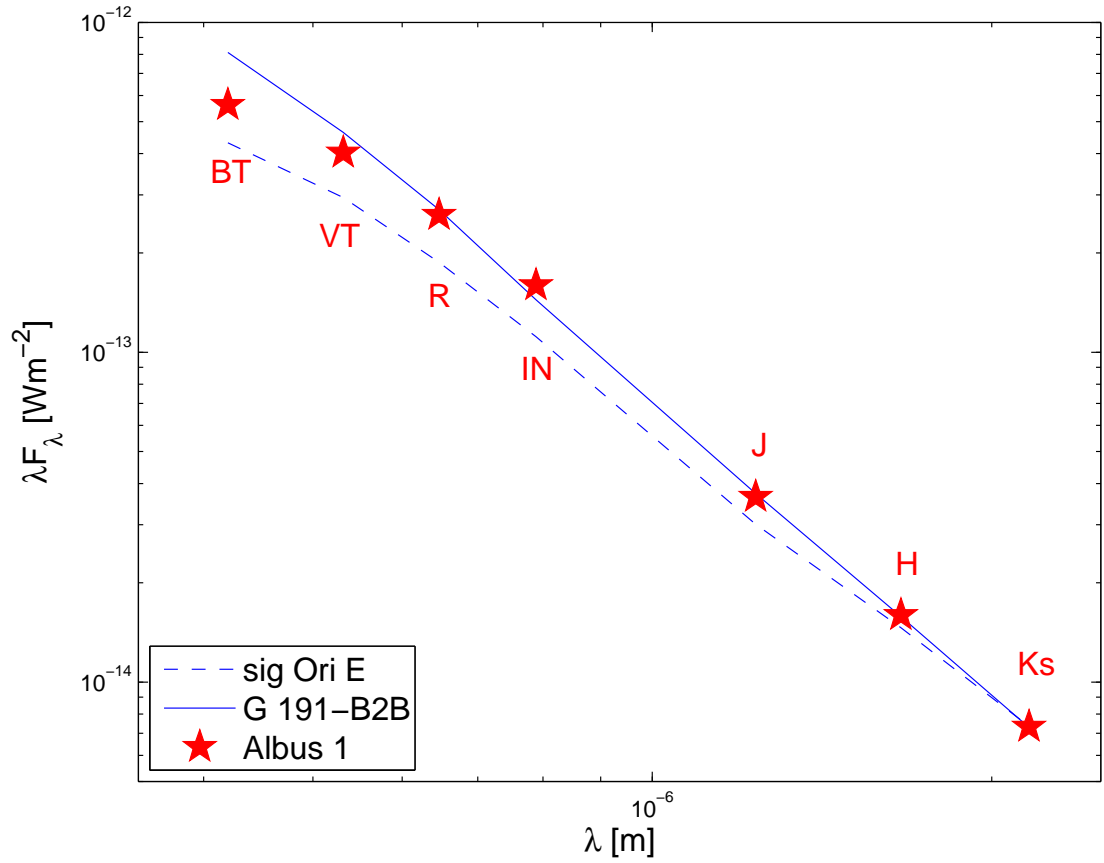}
\caption{Spectral energy distributions of Albus~1, the DA1 white dwarf
G~191--B2B, and the B2Vp star $\sigma$~Ori~E (shifted to an heliocentric
distance of 0.5\,kpc). 
The seven passbands ($B_T V_T R I_N J H K_{\rm s}$) are~indicated.
\label{sed}}
\end{figure}

Data in Table~\ref{WDs} is represented in Fig.~\ref{colorcolor}.
Albus~1 is located in the color-color diagram very close to the well known
white dwarfs G~191--B2B (DA1) and GJ~433.1 (DA3), at $V_T-J \sim$
--0.8\,mag, $J-K_{\rm s} \sim$ --0.2\,mag.
The resemblance between the spectral energy distributions of Albus~1 and
G~191--B2B (``the best studied of all hot white dwarfs''; Barstow et al. 2003),
shown in Fig.~\ref{sed}, is evident. 
Both of them have the same $K_{\rm s}$ magnitude within the error bars (Albus~1:
$K_{\rm s}$ = 12.76$\pm$0.03\,mag; G~191--B2B: $K_{\rm s}$ =
12.76$\pm$0.02\,mag), but G~191--B2B is 0.40$\pm$0.11\,mag brighter in $B_T$.
It leads to tentatively classify Albus~1 as an early DA white dwarf slightly
cooler than G~191--B2B and, therefore, slightly closer to the Sun.
The stellar common proper-motion companion of G~191--B2B has an accurate
parallax determination by {\em Hipparcos} at $d$ = 46$\pm$4\,pc.
Hence, Albus~1 could be located at about 40\,pc, which would explain its
appreciable proper-motion.
The probability of Albus~1 being a more distant blue subdwarf is smaller (see
Table~\ref{WDs}).
From the blue $J-K_{\rm s}$ color in Fig.~\ref{colorcolor}, it is deduced,
besides, that Albus~1 has no main sequence close companion or forms part of a
cataclysmic variable~system. 

As shown by Salim \& Gould (2002), an optical-infrared reduced proper motion
digram (e.g. $V + 5 \log{\mu}$ vs. $V-J$) can be used to classify stars even if
no parallax information is available.  
In particular, white dwarfs and subdwarf stars are easily distinguished from
main sequence stars as they are several magnitudes dimmer at the same color. 
The position of Albus~1 in the reduced proper motion diagram in fig.~4 in Gould
\& Morgan (2003; $V_T + 5 \log{\mu} =$ 3.2$\pm$0.6\,mag, $V_T-J$ =
--0.76$\pm$0.14\,mag) agrees with this requirement.

\section{Conclusions}
\label{conclusions}

Of the 30 white dwarfs and blue subdwarfs listed in Table~\ref{WDs}, only
thirteen have Tycho-2 $V_T$ magnitudes brighter than 12.0\,mag.
Albus~1, with $V_T$ = 11.80$\pm$0.14\,mag, is included in this group.
Accounting for the three brightest known white dwarfs not in the Table, and
discarding the close binary systems BL~Psc~AB, V841~Ara~AB, V3885~Sgr~AB, and
BD+28~4211~AB, whose spectral energy distributions are affected by the main
sequence close companions, then Albus~1 is the 12th brightest white dwarf
yet known. 
Since six of the white dwarfs brighter than it are in multiple systems (the
binary status of GJ~127.1~AB claimed by Gill \& Kaptein 1896 is, however, not
confirmed), then Albus~1 would be the sixth brightest isolated white dwarf,
after the long-time known Feige~34, L~145--141, BD--07~3632, and HD~340611
(Luyten 1949; Eggen \& Greenstein 1965) and the very hot white dwarf and extreme
ultraviolet source RE~J2214--49 (Holberg et~al.~1993).

Albus~1, although located in the southern hemisphere, is visible from the most
important northern observatories.
This fact, together with its brightness, makes our blue source an appropiate
candidate spectrophotometric standard provided that its white dwarf or hot
subdwarf nature is spectroscopically confirmed. 
Our serendipitous detection has also shown that the $V_T-K_{\rm s}$ color is a
good and simple discriminator to look for very blue, relatively bright objects.
A search for new very bright white dwarf candidates using Tycho-2 and 2MASS
is currently~ongoing.

\acknowledgments

This research has made use of the Spanish Virtual Observatory supported 
from the Spanish Ministerio de Educaci\'on y Ciencia through grants
AyA2005--04286 and AyA2005--24102--E of the Plan Nacional de Astronom\'{\i}a y
Astrof\'{\i}sica.  


\clearpage

\begin{deluxetable}{lccc}
\tabletypesize{\scriptsize}
\tablecaption{Basic data of Albus~1.\label{albus1}} 
\tablewidth{0pt}
\tablehead{
	 	& 			& \colhead{Unit}& \colhead{Ref.$^a$}}
\startdata
Name		& Albus~1		&  		& 1	\\	  
WD number	& WD~0604--203		&  		& 1	\\	  
$\alpha$ (J2000)& 06 06 13.39		&  		& 2	\\	  
$\delta$ (J2000)& --20 21 07.3		&  		& 2	\\	  
$\mu_\alpha \cos{\delta}$& +7$\pm$3	& mas\,a$^{-1}$ & 2	\\	  
$\mu_\delta$	& --18$\pm$3		& mas\,a$^{-1}$ & 2	\\	  
$B_T$ 		& 11.75$\pm$0.07	& mag 		& 2	\\	  
$V_T$ 		& 11.80$\pm$0.14	& mag 		& 2	\\	  
$R$ 		& 11.84			& mag 		& 3	\\	  
$I_N$ 		& 11.90			& mag 		& 3	\\	  
$J$ 		& 12.56$\pm$0.02	& mag 		& 4	\\	  
$H$ 		& 12.66$\pm$0.03	& mag 		& 4	\\	  
$K_{\rm s}$ 	& 12.76$\pm$0.03	& mag 		& 4	\\	  
Sp. type	& DA?			&  		& 1	\\	  
$d$		& $\sim$40?		& pc 		& 1	\\	  
\enddata
\tablenotetext{a}{References: 
1,~this work;
2,~Tycho-2 catalogue;
3,~USNO-B1;
4,~2MASS catalogue.}  
\end{deluxetable}

\begin{deluxetable}{llcccc}
\tabletypesize{\scriptsize}
\tablecaption{The brightest white dwarfs and hot subdwarfs in the Tycho-2
catalogue.\label{WDs}}  
\tablewidth{0pt}
\tablehead{
\colhead{WD} & 
\colhead{Alternative} & 
\colhead{Sp.} & 
\colhead{$V_T$} &
\colhead{$J$} &
\colhead{$K_{\rm s}$} \\
\colhead{number} & 
\colhead{name} & 
\colhead{type} & 
\colhead{(mag)} &
\colhead{(mag)} &
\colhead{(mag)} }
\startdata
\object{WD~0041+092}	& \object{BL~Psc}~AB		& DA2+K0IV	&  10.18$\pm$0.04 	&  8.45$\pm$0.03&   7.80$\pm$0.03	\\ 
\object{WD~0046+051}	& \object{van~Maanen's~star}	& DZ7		& 12.559$\pm$0.018 	& 11.69$\pm$0.02&  11.50$\pm$0.02	\\ 
\object{WD~0148+467}	& \object{GJ~3121}		& DA3.5		& 12.552$\pm$0.010 	& 12.77$\pm$0.02&  12.85$\pm$0.03	\\ 
\object{WD~0227+050}	& \object{GJ~100.1}, Feige~22	& DA2.5		& 12.848$\pm$0.019 	& 13.28$\pm$0.03&  13.42$\pm$0.02	\\ 
\object{WD~0232+035}	& \object{FS~Cet}~AB, Feige~24	& DA+M1V	&   12.6$\pm$0.2 	& 11.26$\pm$0.03& 10.557$\pm$0.019	\\ 
\object{WD~0310--688}	& \object{GJ~127.1~A}$^a$	& DA3		&  11.15$\pm$0.07 	& 11.76$\pm$0.02&  11.86$\pm$0.02	\\ 
\object{WD~0426+588}	& \object{GJ~169.1~B}$^b$	& DQ7		&  12.13$\pm$0.18 	&  6.62$\pm$0.02&   5.72$\pm$0.02	\\ 
\object{WD~0501+527}	& \object{G~191--B2B}		& DA1		&  11.65$\pm$0.17 	& 12.54$\pm$0.02&  12.76$\pm$0.02	\\ 
\bf{WD~0604--203}	& \bf{Albus~1}			& \bf{DA?}  &  \bf{11.80$\pm$0.14}  & \bf{12.56$\pm$0.02}& \bf{12.76$\pm$0.03}	\\ 
\object{WD~0621--376}	& \object{RE~J0623--374}	& DA1		&   12.3$\pm$0.2 	& 12.85$\pm$0.03&  13.09$\pm$0.03	\\ 
\object{WD~0644+375}	& \object{GJ~246}, He~3		& DA2.5		&   12.2$\pm$0.2 	&  12.7$\pm$0.3 &   12.8$\pm$0.3	\\ 
\object{WD~0713+584}	& \object{GJ~3435}, GD~294	& sdB		&  12.03$\pm$0.16 	& 11.77$\pm$0.02& 11.721$\pm$0.018	\\ 
\object{WD~1036+433}	& \object{GJ~398.2}, Feige~34	& DA0:		&  11.10$\pm$0.07 	& 11.64$\pm$0.02& 11.540$\pm$0.019	\\ 
\object{WD~1134+300}	& \object{GJ~433.1}		& DA3		&  12.14$\pm$0.19 	& 12.99$\pm$0.02&  13.18$\pm$0.03	\\ 
\object{WD~1142--645}	& \object{GJ~440}, L~145--141	& DQ6		&  11.34$\pm$0.09 	& 11.19$\pm$0.02&  11.10$\pm$0.03	\\ 
\object{WD~1314+293}	& \object{HZ~43}~AB$^c$		& DA1+M3.5Ve	& 12.667$\pm$0.009 	&10.373$\pm$0.019&  9.56$\pm$0.02	\\ 
\object{WD~1327-083}	& \object{BD--07~3632}$^d$	& DA4		&  11.77$\pm$0.18 	& 12.62$\pm$0.04&  12.74$\pm$0.05	\\ 
\object{WD~1337+705}	& \object{G~238--44}		& DA3		& 12.839$\pm$0.010 	& 13.25$\pm$0.02&  13.45$\pm$0.04	\\ 
\object{WD~1620--391}	& \object{HD~147513~B}$^e$	& DA2		&  11.00$\pm$0.08 	& 11.58$\pm$0.02&  11.77$\pm$0.02	\\ 
\object{WD~1634--573}	& \object{V841~Ara}~AB$^f$	& DOZ1+K0V	&  11.26$\pm$0.07 	&  7.12$\pm$0.02&   6.57$\pm$0.03	\\ 
\object{WD~1647+591}	& \object{DN~Dra}, G~226--29$^g$& DAV4.7	&  12.00$\pm$0.16 	& 12.42$\pm$0.02&  12.52$\pm$0.03	\\ 
\object{WD~1917--077}	& \object{GJ~754.1~A}$^h$	& DBQA5		&   13.1$\pm$0.3 	& 12.35$\pm$0.03&  12.42$\pm$0.03	\\ 
\object{WD~1944--421}	& \object{V3885~Sgr}~AB$^i$	& DB:p+M:V	&  10.33$\pm$0.05 	&  9.96$\pm$0.03&   9.62$\pm$0.02	\\ 
\object{WD~2007--303}	& \object{GJ~2147}		& DA4		&   12.9$\pm$0.3 	& 12.58$\pm$0.02&  12.70$\pm$0.03	\\ 
\object{WD~2032+248}	& \object{HD~340611}		& DA2.5		&  11.55$\pm$0.10 	& 12.04$\pm$0.03&  12.19$\pm$0.03	\\ 
\object{WD~2039--202}	& \object{GJ~7991.1}		& DA2.5		&   13.1$\pm$0.3 	& 12.82$\pm$0.03&  13.00$\pm$0.03	\\ 
\object{WD~2117+539}	& \object{V2151~Cyg}		& DA3.5		&   12.7$\pm$0.3 	& 12.68$\pm$0.02&  12.85$\pm$0.04	\\ 
\object{WD~2148+286}	& \object{BD+28~4211}~AB$^j$	& sdO:p+G:	&  10.53$\pm$0.05 	& 11.28$\pm$0.03&  11.56$\pm$0.03	\\ 
\object{WD~2149+021}	& \object{GJ~838.4}		& DA3		&   13.2$\pm$0.4 	& 13.20$\pm$0.02&  13.39$\pm$0.04	\\ 
\object{WD~2211--495}	& \object{RE~J2214--49}		& DA.76		&  11.52$\pm$0.08 	& 12.44$\pm$0.03&  12.64$\pm$0.03	\\ 
\enddata
\tablenotetext{a}{The status of the hypothetical companion \object{GJ~127.1~B}
is~unknown.}  
\tablenotetext{b}{The 2MASS pipeline did not identify the white dwarf, which
is visible in $JHK_{\rm}$ at $\sim$8.0\,arcsec from \object{GJ~169.1~A}.}  
\tablenotetext{c}{Neither Tycho-2 nor 2MASS resolved the HZ~43~AB system 
(L~1409--4~AB), whose components are separated by 2.23$\pm$0.03\,arcsec 
(McAlister et~al.~1996).}  
\tablenotetext{d}{Common proper motion companion of the low-mass star 
\object{GJ~514.1} (LHS~353;~M4.5V).}
\tablenotetext{e}{Common proper motion companion of the exoplanet-harbour star 
\object{HD~147513} (G5V; Desidera \& Barbieri~2007).}
\tablenotetext{f}{2MASS did not resolve the V841~Ara~AB system 
(HD~149499~AB), whose components are separated by 1.320\,arcsec (Tycho-2).}  
\tablenotetext{g}{ZZ~Ceti-type pulsating white~dwarf.}
\tablenotetext{h}{Common proper motion companion of the low-mass star 
\object{GJ~754.1~B} (L~923--22;~M3.5V).}
\tablenotetext{i}{UX~UMa nova-like cataclysmic variable in which the secondary 
is a $\sim$0.5\,$M_\odot$ red dwarf that fills its Roche lobe (orbital period: 
$P \approx$ 0.207\,d; Ribeiro \& Diaz~2007).} 
\tablenotetext{j}{The binary status of BD+28~4211 has not been confirmed.}
\end{deluxetable}

\clearpage

\clearpage

\clearpage

\clearpage

\end{document}